\documentclass[11pt,a4paper]{article}
\usepackage[margin=2cm]{geometry}
\usepackage[T1]{fontenc}
\usepackage[utf8]{inputenc}
\usepackage[noadjust]{cite}
\usepackage{lmodern}
\usepackage{amsthm}
\usepackage{thmtools}
\makeatletter
\@ifundefined{newcounteralias}{}{%
  \renewcommand\thmt@autorefsetup{%
    \@xa\def\csname\thmt@envname autorefname\@xa\endcsname
      \@xa{\thmt@thmname}%
  }%
}
\makeatother

\usepackage{thm-autoref}
\usepackage[shortlabels]{enumitem}
\usepackage[center]{caption}
\usepackage{latexsym}
\usepackage{amsfonts}
\usepackage{graphicx}
\usepackage{mathrsfs}
\usepackage[normalem]{ulem}
\usepackage{amssymb, mathdots, mathtools, xkeyval, tikz, tkz-euclide, adjustbox, float, array, subcaption, listings}
\usepackage{comment}
\usepackage{xspace}
\usepackage[colorlinks=true,
linkcolor =blue,
citecolor=blue,
urlcolor=blue]{hyperref}
\usepackage[capitalise]{cleveref}

\theoremstyle{plain}
\newtheorem{problem}{Problem}[section]
\newtheorem{theorem}{Theorem}[section]
\newtheorem{lemma}[theorem]{Lemma}
\newtheorem{proposition}[theorem]{Proposition}

\newtheorem{claim}{Claim}[theorem]
\newenvironment{claimproof}{\noindent\textit{Proof of Claim \theclaim:}}{\hfill$\lrcorner$\\} 

\crefname{claim}{Claim}{Claims}

\newcommand{\MWIS}{\textsc{MWIS}\xspace}
\newcommand{\NP}{\textsf{NP}\xspace}
\newcommand{\Oh}{\mathcal{O}}

\newcommand{\weight}{\textsf{w}}
\newcommand{\Tab}{\mathsf{Tab}}
\let\smallfontsize\small
\renewcommand{\small}{\ifmmode\mathsf{small}\else\smallfontsize\fi}
\newcommand{\Small}{\mathsf{Small}}
\renewcommand{\phi}{\varphi}

\newcommand{\cA}{\mathcal{A}}

\newcommand{\cF}{\mathcal{F}}

\newcommand{\cH}{\mathcal{H}}

\newcommand{\cQ}{\mathcal{Q}}
\newcommand{\cS}{\mathcal{S}}

\usepackage{todonotes}

\title{Max Weight Independent Set in graphs\\
with no long claws nor large induced bicliques}

\author{Maria Chudnovsky\thanks{Princeton University. Supported by NSF Grant DMS-2348219, NSF Grant CCF-2505100, AFOSR grant FA9550-25-1-0275 and a Guggenheim Fellowship}
\and Paweł Rzążewski\thanks{Warsaw University of Technology. Supported by the National Science Centre grant 2024/54/E/ST6/00094.}}
\date{}

\hypersetup{hypertexnames=false}
\begin{document}
	\begin{titlepage}
	\date{}
	\maketitle

\begin{abstract}
    The complexity of \textsc{Max Weight Independent Set} (\textsc{MWIS}) in graphs excluding a fixed subdivided claw is one of the central open problems in algorithmic graph theory.
We show that \textsc{MWIS} can be solved in polynomial time in graphs excluding a fixed subdivided claw and a fixed biclique as an induced subgraph.
This strengthens a result of Abrishami, Chudnovsky, Dibek, Pilipczuk, and Rzążewski [ACM TALG 2026] who showed an analogous result, but under a stronger assumption that some biclique is excluded as a subgraph.
\end{abstract}

	\thispagestyle{empty}

  \bigskip
  \bigskip
  \bigskip
\tableofcontents
\end{titlepage}
\thispagestyle{empty}
\newpage\setcounter{page}{1}

\section{Introduction}

\textsc{Maximum Weight Independent Set} (\MWIS) is one of the fundamental problems in algorithmic graph theory.
Given a graph with nonnegative vertex weights, the task is to find a set of pairwise nonadjacent vertices (i.e., an independent set) of maximum total weight.
The unweighted version of this problem is equivalent, by graph complementation, to \textsc{Clique}, one of Karp's 21 \NP-hard problems~\cite{Karp1972}.
Nevertheless, \MWIS admits polynomial-time algorithms in many important graph classes, including perfect graphs~\cite{DBLP:journals/combinatorica/GrotschelLS81}, claw-free graphs~\cite{MINTY1980284,SBIHI198053,clawCubic}, and circle graphs (these are  the intersection graphs of chords of a circle)~\cite{DBLP:conf/isaac/Valiente03}.

We are interested in the complexity of \MWIS in graph classes defined by forbidden induced subgraphs.
For two graphs $F$ and $G$, the
 graph  $G$ is \emph{$F$-free} if $G$ contains no induced subgraph isomorphic to $F$. More generally, $G$  is \emph{$(F_1,\ldots,F_k)$-free} if it is $F_i$-free for every $i$.
For $t\geq1$, let $P_t$ denote the path on $t$ vertices, i.e., of length $t-1$.
For $a,b,c\geq1$, the \emph{subdivided claw} $S_{a,b,c}$ is obtained from three paths of lengths $a,b,c$ by identifying one endpoint of each path into a single vertex.
In particular, $S_{1,1,1}=K_{1,3}$ is the three-leaf star, also known as the \emph{claw}.
Let $\cF$ be the family of all graphs  $G$ such that each component of $G$ 
is a subcubic tree with at most three leaves, or equivalently, a path or a subdivided claw.

Alekseev~\cite{alekseev1982effect} showed that, for every fixed $F\notin\cF$, even the problem of finding an independent set of maximum size (this is a special case of MWIS with all equal vertx weights)  remains \NP-hard in $F$-free graphs.
For $F\in\cF$, polynomial-time algorithms are known only in special cases.
Notable examples include $P_5$-free graphs, by Lokshtanov, Vatshelle, and Villanger~\cite{LokshtanovVV14}; $P_6$-free graphs, by Grzesik, Klimo\v{s}ov\'a, Pilipczuk, and Pilipczuk~\cite{GrzesikKPP22}; and fork-free graphs, by Lozin and Milani\v{c}~\cite{DBLP:journals/jda/LozinM08}, where the fork is $S_{2,1,1}$.
There are also nontrivial tractable cases with disconnected forbidden graphs~\cite{BM18-lclaw,DBLP:journals/tcs/BrandstadtLM25}.

Interestingly, for every $F \in \cF$, \MWIS in $F$-free graphs can be solved in \emph{quasipolynomial} time.
This was first established by Gartland and Lokshtanov~\cite{GartlandL20} (see also~\cite{PilipczukPR21}) for the case that $F$ is a path,
and then extended to general $F \in \cF$ by Gartland, Lokshtanov, Masa\v{r}\'ik, Pilipczuk, Pilipczuk, and Rz\k{a}\.{z}ewski~\cite{DBLP:conf/stoc/GartlandLMPPR24}.
Whether these cases all admit polynomial-time algorithms remains wide open.

This motivates studying $F$-free graphs under additional restrictions.
For example, polynomial-time algorithms for \MWIS are known for graphs that exclude $P_7$ and some small graph~\cite{BM18,p72,p73}.
Recently, \MWIS was shown to be polynomial-time-solvable in $P_7$-free graphs of bounded clique number by Chudnovsky, Czy\.zewska, Kluk, Pilipczuk, and Rz\k{a}\.{z}ewski~\cite{chudnovsky_et_al:LIPIcs.ISAAC.2025.20}.
Another direction is to impose sparsity conditions.
For every fixed $F\in\cF$ and integer $\Delta$, Abrishami, Chudnovsky, Dibek, and Rz\k{a}\.{z}ewski~\cite{ACDR21} gave a polynomial-time algorithm for \MWIS in $F$-free graphs of maximum degree at most $\Delta$.
This was strengthened by Abrishami, Chudnovsky, Dibek, Pilipczuk, and Rz\k{a}\.{z}ewski~\cite{10.1145/3844498}, who replaced the degree bound with the assumption that a fixed biclique is excluded as a subgraph, not necessarily induced.
In this paper we further generalize these results by requiring only that a fixed biclique be excluded as an induced subgraph.

\begin{restatable}{theorem}{thmmain}
    \label{thm:main}
    For every fixed $s \geq 1$ and $F \in \cF$, there is a polynomial-time algorithm which, given an $(F,K_{s,s})$-free graph $G$ and a weight function $\weight:V(G)\to\mathbb{R}_{\geq 0}$, returns a maximum-weight independent set in $(G,\weight)$.
\end{restatable}

We emphasize that, by the celebrated K\"ovari--S\'os--Tur\'an theorem~\cite{MR65617}, excluding a fixed biclique as a subgraph forces the number of edges to be subquadratic,
and in $F$-free graphs, the number of edges becomes even linear the result of Kuhn and Osthus~\cite{KuhnOsthus04}.
Our setting permits quadratically many edges, so the algorithm must also accommodate dense graphs.

Our proof combines the algorithmic use of extended strip decompositions developed by Chudnovsky, Pilipczuk, Pilipczuk, and Thomass\'e~\cite{DBLP:journals/siamcomp/ChudnovskyPPT24} with the boundary formulation of Abrishami et al.~\cite{10.1145/3844498}.
An extended strip decomposition allows solutions on overlapping pieces, called \emph{particles}, to be assembled in polynomial time.
The key is to control the independence number of separators and boundaries, which may contain arbitrarily many vertices.

Bounding the independence number is sufficient for our purposes: every independent set contains only a bounded number of vertices from a set of bounded independence number, so all possible intersections of a solution with such a set can be enumerated in polynomial time.
Combining structural results of Hajebi and Spirkl~\cite{DBLP:journals/corr/abs-2605-01223} with those of Bourneuf, Masa\v{r}\'ikov\'a, Nadara, and Pilipczuk~\cite{DBLP:conf/mfcs/BourneufMNP25}, we show how to efficiently find, in a graph excluding a fixed subdivided claw and a fixed biclique, a set of bounded independence number whose deletion leaves a graph with an extended strip decomposition in which every particle contains at most half of the vertices from a prescribed set.
We use such decompositions recursively: from each particle, we remove a set of bounded independence number and recurse on the components of the remaining graph.
An appropriate choice of the prescribed set ensures that the recursion depth is logarithmic.
For each such component, we consider every possible intersection of a solution with its \emph{boundary}, which is a certain set of neighbors of the component, and compute the best way to extend it inside the component.
A new ingredient is a bound on the independence number of these boundaries, which guarantees that only polynomially many such intersections need to be considered.
As each vertex belongs to only a bounded number of the components considered in a single recursive call, the total running time is polynomial.
Finally, if $F$ has several components, we again use the machinery of Hajebi and Spirkl~\cite{DBLP:journals/corr/abs-2605-01223} to find a set of bounded independence number that intersects every induced copy of a sufficiently long subdivided claw; deleting this set reduces the problem to graphs excluding a single subdivided claw.


\section{Preliminaries}

\paragraph{Weights.}
Let $G$ be a graph and let $\weight : V(G) \to \mathbb{R}_{\geq 0}$ be a weight function on the vertices of $G$.
A vertex-weighted graph is a pair $(G,\weight)$.
We also use $\weight$ for its restriction to any induced subgraph of $G$.
For a set $X \subseteq V(G)$, we define $\weight(X) := \sum_{v \in X} \weight(v)$,
and we write $\weight(G) := \weight(V(G))$.
Arithmetic operations and comparisons of weights take constant time.

In the paper we will use two types of weight functions: one coming from the input of \MWIS, 
and another indicating some ``important'' vertices in the graph.
To avoid confusion, the second function will be called a \emph{mass function} and typically denoted by $\mu$.
Usually, the mass will be defined for a subset $U \subseteq V(G)$ as $\mu(v) = 1$ if $v \in U$ and $\mu(v) = 0$ otherwise. We will call such a function \emph{indicator mass function of $U$}.

\paragraph{Graph theory.}
For two graphs $G_1$ and $G_2$, we denote by $G_1 + G_2$ the disjoint union of $G_1$ and $G_2$.
For $X\subseteq V(G)$, let $N_G(X)$ be the set of neighbors of $X$, i.e., vertices outside $X$ with a neighbor in $X$, and write $N_G[X]=X\cup N_G(X)$.
Two sets are \emph{anticomplete} if they are disjoint and no edge joins them.
For a graph $G$, by $\alpha(G)$ we denote the independence number of $G$, i.e., the size of a maximum independent set in $G$.
For a vertex-weighted graph $(G,\weight)$, by $\alpha(G,\weight)$ we denote the maximum weight of an independent set in $G$ with respect to $\weight$.
If $G$ is clear from the context, for $X \subseteq V(G)$ we will write $\alpha(X)$ and $\alpha(X,\weight)$ instead of $\alpha(G[X])$ and $\alpha(G[X],\weight)$, respectively. 

\paragraph{Paths in graphs.}
Let $G$ be a graph and let $A,B$ be disjoint subsets of $V(G)$.
By $A$--$B$-path we mean a path in $G$ with one endpoint in $A$, the other endpoint in $B$, and all internal vertices in $V(G) \setminus (A \cup B)$.
An \emph{$A$--$B$-separator} is a set $Y \subseteq V(G)\setminus(A\cup B)$ that intersects all $A$--$B$-paths in $G$.

\begin{figure}
\includegraphics[scale=1]{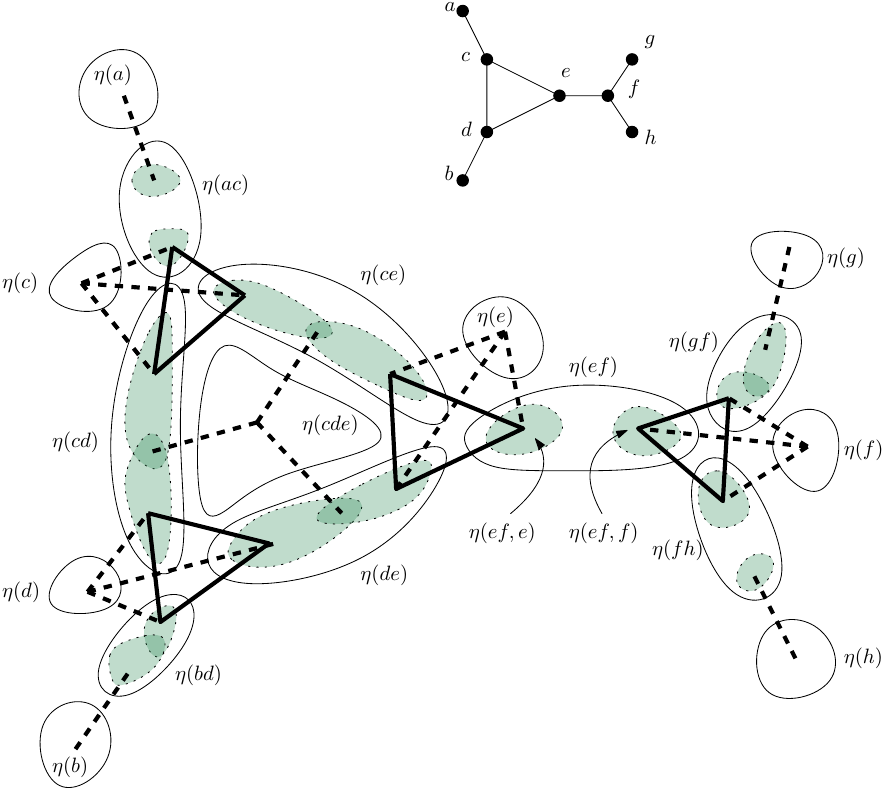}
\caption{A graph $H$ (top) and a schematic view of an extended strip decomposition $(H,\eta)$ of a graph $G$.
Green areas depict the sets $\eta(xy,y)$.
Solid line indicates that all edges between two sets exist. Dashed line denotes that edges might exist.
Picture from~\cite{DBLP:conf/stoc/GartlandLMPPR24}, courtesy of the authors.}
\label{fig:esd}
\end{figure}

\paragraph{Extended strip decompositions.}
We follow the notation of~\cite{ICALP-qptas,DBLP:conf/stoc/GartlandLMPPR24}.
A \emph{triangle} in a graph is a set of three distinct, pairwise adjacent vertices.
For a graph $H$, by $T(H)$ we denote the set of triangles in $H$.
We write $xyz$ for a triangle $\{x,y,z\}$.
An \emph{extended strip decomposition} of a graph $G$ is a pair $(H, \eta)$ that consists of:
\begin{itemize}
\item a simple graph $H$ called the \emph{host},
\item a \emph{vertex atom} $\eta(x) \subseteq V(G)$ for every $x \in V(H)$,
\item an \emph{edge atom} $\eta(xy) \subseteq V(G)$ for every $xy \in E(H)$, and its subsets $\eta(xy,x),\eta(xy,y) \subseteq \eta(xy)$ called \emph{interfaces}, and
\item a \emph{triangle atom} $\eta(xyz) \subseteq V(G)$  for every $xyz \in T(H)$,
\end{itemize}
which have the following properties (empty sets are allowed):
\begin{enumerate}
\item The family $\{\eta(o)\mid o \in V(H)\cup E(H) \cup T(H)\}$ is a partition of $V(G)$.
\item For every $x \in V(H)$ and every distinct $y,z \in N_H(x)$, the set $\eta(xy,x)$ is complete to $\eta(xz,x)$ (i.e., all possible edges between these two sets exist).
\item Every $uv \in E(G)$ is contained in one of the atoms $\eta(o)$ for $o \in V(H) \cup E(H)\cup T(H)$, or is as follows:
\begin{itemize}
\item $u \in \eta(xy,x), v\in \eta(xz,x)$ for some $x \in V(H)$ and $y,z \in N_H(x)$, or
\item $u \in \eta(xy,x), v\in \eta(x)$ for some $xy \in E(H)$, or
\item $u \in \eta(xyz)$ and $v\in \eta(xy,x) \cap \eta(xy,y)$ for some $xyz \in T(H)$. 
\end{itemize}
\end{enumerate}
A schematic picture of an extended strip decomposition of a graph is shown in \Cref{fig:esd}.

An extended strip decomposition $(H,\eta)$ is \emph{rigid} if for every $xy \in E(H)$, the sets $\eta(xy,x)$ and
$\eta(xy,y)$ are nonempty, and for every isolated $x \in V(H)$, the vertex atom $\eta(x)$ is nonempty.

For an extended strip decomposition $(H,\eta)$ of a graph $G$, we identify five \emph{types}
of \emph{particles}.
\begin{align*}
\textrm{vertex particle:} &\quad A_{x} := \eta(x) \text{ for each } x \in V(H),\\
\textrm{edge interior particle:} &\quad A_{xy}^{\perp} := \eta(xy) \setminus (\eta(xy,x) \cup \eta(xy,y)) \text{ for each } xy \in E(H),\\
\textrm{half-edge particle:} &\quad A_{xy}^{x} :=  \eta(x) \cup \eta(xy) \setminus \eta(xy,y) \text{ for each } xy \in E(H),\\
\textrm{full edge particle:} &\quad A_{xy}^{xy} := \eta(x) \cup \eta(y) \cup \eta(xy) \cup \bigcup_{z \mid xyz \in T(H)} \eta(xyz) \text{ for each } xy \in E(H),\\
\textrm{triangle particle:} &\quad A_{xyz} := \eta(xyz) \text{ for each } xyz \in T(H).
\end{align*}

Given a mass function $\mu$ on $V(G)$, a decomposition of an induced
subgraph of $G$ is \emph{balanced for $\mu$} if every particle $A$ satisfies
$\mu(A)\leq\mu(G)/2$.

Let us conclude with listing some straightforward properties of extended strip decompositions that will be used in the paper. The first two follow directly from the definition of extended strip decompositions, and the last one is shown in~\cite[Observation 5]{ICALP-qptas}.

\begin{proposition}\label{lem:restriction}
    The following hold.
    \begin{enumerate}[label=(\arabic*),ref=(\arabic*)]
        \item\label{it:restriction-induced} Let $G$ be a graph, let $(H,\eta)$ be an extended strip decomposition of $G$, and let $X \subseteq V(G)$.
        By restricting $\eta$ to $X$, we obtain an extended strip decomposition of $G[X]$, with every particle replaced by its intersection with $X$.
        \item\label{it:restriction-union} Let $G_1,G_2$ be two graphs on disjoint vertex sets, let $(H_1,\eta_1)$ and $(H_2,\eta_2)$ be extended strip decompositions of $G_1$ and $G_2$, respectively.
        Then, by taking the disjoint union of $H_1$ and $H_2$ and the union of $\eta_1$ and $\eta_2$, we obtain an extended strip decomposition of $G_1 + G_2$.
        \item\label{it:restriction-rigid} If $(H,\eta)$ is a rigid extended strip decomposition of $G$, then $|V(H)| \leq 2|V(G)|$ and $|E(H)| \leq |V(G)|$.
    \end{enumerate}
\end{proposition}

The main algorithmic application of extended strip decompositions is that, given an extended strip decomposition $(H,\eta)$ of $(G,\weight)$, one can reduce computing $\alpha(G,\weight)$ to solving \MWIS in particles of $(H,\eta)$.

\begin{theorem}[{Chudnovsky, Pilipczuk, Pilipczuk, and Thomass\'e~\cite[Section 3.3]{DBLP:journals/siamcomp/ChudnovskyPPT24}}]\label{thm:assembly}
Given a weighted graph $(G,\weight)$, an extended strip decomposition $(H,\eta)$ of $G$,
and the value $\alpha(A,\weight)$ for every particle $A$, one can compute $\alpha(G,\weight)$
in time polynomial in $|V(G)|+|V(H)|$.
\end{theorem}


\section{Auxiliary tools}


In this section we develop a number of tools that will be used in the proof of \cref{thm:main}.

\subsection{Hitting sets of small independence number}

For a graph $G$ and an integer $b$, a \emph{$b$-system in $G$} is a family of nonempty vertex sets of $G$,
each of size at most $b$; the family itself may be empty.
A set \emph{hits} a family if it intersects every member. Note that an empty set system is hit by the empty set.
Hajebi and Spirkl~\cite{DBLP:journals/corr/abs-2605-01223} proved that in graphs excluding a fixed induced biclique, $b$-systems admit a certain Erd\H{o}s--P\'osa-type property: either we can find many pairwise anticomplete members of the system, or a hitting set of small independence number.

\begin{lemma}[{Hajebi and Spirkl~\cite[Lemma 6.2]{DBLP:journals/corr/abs-2605-01223}}]\label{lem:hit-vs-pack}
For $b,s,q \geq 1$,
there is a computable constant $c_{\ref{lem:hit-vs-pack}}=c_{\ref{lem:hit-vs-pack}}(b,s,q)\in\mathbb{N}$ such that, 
for every $K_{s,s}$-free graph $G$ and every $b$-system $\cS$ in $G$, one of the following holds:
\begin{enumerate}
    \item there is a hitting set $X\subseteq V(G)$ for $\cS$ with $\alpha(X) \leq c_{\ref{lem:hit-vs-pack}}$; or
    \item there are $q$ pairwise anticomplete members of $\cS$.
\end{enumerate}
\end{lemma}

\begin{lemma}[{Hajebi and Spirkl~\cite[Theorem 3.1]{DBLP:journals/corr/abs-2605-01223}}]\label{lem:hit-vs-far}
For every $b,s,t \geq 1$,
there are computable constants $c_{\ref{lem:hit-vs-far}}=c_{\ref{lem:hit-vs-far}}(b,s,t)\in\mathbb{N}$ and
$d_{\ref{lem:hit-vs-far}}=d_{\ref{lem:hit-vs-far}}(b,s,t)\in\mathbb{N}$ such that, 
for every $(S_{t,t,t},K_{s,s})$-free graph $G$ and every $b$-system $\cS$ in $G$, one of the following holds:
\begin{enumerate}
    \item there is a hitting set $X\subseteq V(G)$ for $\cS$ with $\alpha(X) \leq c_{\ref{lem:hit-vs-far}}$; or
    \item there are anticomplete $S_1,S_2\in\cS$ and an $S_1$--$S_2$-separator $Y$ with $\alpha(Y) \leq d_{\ref{lem:hit-vs-far}}$.
\end{enumerate}
\end{lemma}

\begin{lemma}[Algorithmic \cref{lem:hit-vs-pack}]\label{lem:packing}
For fixed $b,s,q\geq1$, there is a computable constant $c_{\ref{lem:packing}}=c_{\ref{lem:packing}}(b,s,q)$ and a polynomial-time algorithm which, given a  $b$-system $\cS$ in a $K_{s,s}$-free graph $G$, returns either:
\begin{enumerate}
    \item a hitting set $X$ for $\cS$ with $\alpha(X)\leq c_{\ref{lem:packing}}$; or
    \item $q$ pairwise anticomplete members of $\cS$.
\end{enumerate}
\end{lemma}
\begin{proof}
Enumerate all $q$-tuples of members of $\cS$ and return an anticomplete tuple if it
exists. This can be done in polynomial time, as both $b$ and $q$ are fixed.

Otherwise, \cref{lem:hit-vs-pack} guarantees a hitting set of
independence number at most $c_{\ref{lem:hit-vs-pack}}=c_{\ref{lem:hit-vs-pack}}(b,s,q)$.
Without loss of generality we may assume that this hitting set is contained in $\bigcup \cS$.
Solve the rational feasibility program
\[
\begin{aligned}
 0\leq x_v&\leq1 &&\text{ for every } v\in \bigcup \cS,\\
 \sum_{v\in S}x_v&\geq1 &&\text{ for every }  S \in\cS,\\
 \sum_{v\in I}x_v&\leq c_{\ref{lem:hit-vs-pack}}
 &&\text{ for every independent set } I\subseteq \bigcup \cS \text{ with } |I|=b \cdot c_{\ref{lem:hit-vs-pack}} +1.
\end{aligned}
\]
This linear program has a solution:
indeed, the indicator vector of the guaranteed hitting set satisfies every constraint.
Take $X=\{v \mid x_v\geq1/b\}$.

By the second family of constraints, for every $S \in \cS$,
we have $\sum_{v\in S}x_v\geq1$, and hence $X\cap S\neq\emptyset$.
On the other hand, $\alpha(X) \leq b \cdot c_{\ref{lem:hit-vs-pack}}$, as otherwise we would have an independent set $I\subseteq X \subseteq \bigcup \cS$ of size $b \cdot c_{\ref{lem:hit-vs-pack}} +1$, and then $\sum_{v\in I}x_v\geq |I|/b>c_{\ref{lem:hit-vs-pack}}$, violating the last constraint of the program.
Thus we can set $c_{\ref{lem:packing}}(b,s,q)=b\cdot c_{\ref{lem:hit-vs-pack}}(b,s,q)$.

Note that the number of constraints is polynomial in $|V(G)|$, as $b$ and $c_{\ref{lem:hit-vs-pack}}$ are constants.
As linear programs can be solved in polynomial time, the proof is complete.
\end{proof}

\begin{lemma}[Algorithmic \cref{lem:hit-vs-far}]\label{lem:separation}
For every $b,s,t \geq 1$,
there are constants $c_{\ref{lem:separation}}=c_{\ref{lem:separation}}(b,s,t)\in\mathbb{N}$ and
$d_{\ref{lem:separation}}=d_{\ref{lem:separation}}(b,s,t)\in\mathbb{N}$ and a polynomial-time algorithm that, given an $(S_{t,t,t},K_{s,s})$-free graph $G$ and a $b$-system $\cS$ in $G$, returns either:
\begin{enumerate}
    \item a hitting set $X\subseteq V(G)$ for $\cS$ with $\alpha(X) \leq c_{\ref{lem:separation}}$; or
    \item two anticomplete members $S_1,S_2\in\cS$ and an $S_1$--$S_2$-separator $Y$ with $\alpha(Y) \leq d_{\ref{lem:separation}}$.
\end{enumerate}
\end{lemma}
\begin{proof}
    Define the following constants depending only on $b,s,t$:
    \begin{align*}
        q&=c_{\ref{lem:hit-vs-far}}(b,s,t)+1,\\
        c_{\ref{lem:separation}}&=c_{\ref{lem:packing}}(b,s,q),\\
        d_{\ref{lem:separation}}&=c_{\ref{lem:packing}}\bigl(t,s,d_{\ref{lem:hit-vs-far}}+2b+1\bigr).
    \end{align*}
Apply \cref{lem:packing} to $\cS$ with parameters $b,s,q$.
If the call returns the first outcome, i.e., a hitting set with independence number at most $c_{\ref{lem:separation}}$, we return it as the first outcome.

Thus, suppose that the application of \cref{lem:packing} returns a subfamily 
$\cA \subseteq \cS$ consisting of $q$ pairwise anticomplete members of $\cS$.
Every hitting set for $\cA$ has independence number at least $q$, as the $q$ sets in $\cA$ are pairwise anticomplete and we have to pick a vertex from each of them.
Thus \cref{lem:hit-vs-far} applied to $\cA$ guarantees the existence of two members $\widehat{S}_1,\widehat{S}_2$ with a corresponding separator $\widehat{Y}$ of independence number at most $d_{\ref{lem:hit-vs-far}}$.
We are left with finding such a pair and the corresponding separator of independence number at most $d_{\ref{lem:separation}}$ in polynomial time.

For every unordered pair of distinct $S_1,S_2\in\cA$, we proceed as follows.
Define $W=V(G)\setminus(S_1\cup S_2)$ and construct a $t$-system $\cH_{S_1,S_2}$ in $G[W]$ consisting of:
\begin{enumerate}
    \item the sets of internal vertices of all induced $S_1$--$S_2$-paths of length at most $t$; and
    \item all sets $\{v_1,\ldots,v_t\}\subseteq W$ such that $v_0,v_1,\ldots,v_t$ is an induced path for some $v_0\in S_1$.
\end{enumerate}
According to the definition above, we distinguish the members of $\cH_{S_1,S_2}$ of the first and second type, respectively.
Note that every $S_1$--$S_2$-path has at least one internal vertex, as $S_1$ and $S_2$ are anticomplete.
Consequently, every member of $\cH_{S_1,S_2}$ is nonempty and has size at most $t$.

\begin{claim}\label{clm:hitting-to-separator}
Every hitting set $Y\subseteq W$ for $\cH_{S_1,S_2}$ is an $S_1$--$S_2$-separator.
\end{claim}
\begin{claimproof}
Suppose otherwise and let $P$ be a shortest $S_1$--$S_2$-path in $G-Y$.
In particular, $P$ is induced.
If $P$ has length at most $t$, then the set of its internal vertices is a member of $\cH_{S_1,S_2}$ of the first type, contradicting the choice of $Y$.
Otherwise, $t$ consecutive vertices of $P$, starting at the neighbor of the endpoint in $S_1$, form a member of $\cH_{S_1,S_2}$ of the second type, again contradicting the choice of $Y$.
\end{claimproof}

\begin{claim}
For every $S_1,S_2$, the system $\cH_{S_1,S_2}$ contains at most $2b$ pairwise anticomplete members of the second type.
\end{claim}
\begin{claimproof}
Suppose otherwise.
By the pigeonhole principle, three witnessing paths would have a common endpoint in $S_1$, since $|S_1|\leq b$. As, except for the common endpoint, the paths are pairwise anticomplete, they would form an induced $S_{t,t,t}$ in $G$, a contradiction.
\end{claimproof}

For the pair $\widehat{S}_1,\widehat{S}_2$, there are also at most $d_{\ref{lem:hit-vs-far}}$ pairwise anticomplete members of the first type:
the separator $\widehat{Y}$ hits each of them, and choosing one vertex from each intersection gives an independent set in the separator.
Consequently, $\cH_{\widehat{S}_1,\widehat{S}_2}$ has no $d_{\ref{lem:hit-vs-far}}+2b+1$ pairwise anticomplete members.

Apply \cref{lem:packing} to each $\cH_{S_1,S_2}$ in $G[W]$ with parameters
$t,s,d_{\ref{lem:hit-vs-far}}+2b+1$.
Since the second outcome is impossible for the pair $\widehat{S}_1,\widehat{S}_2$, at least one call returns a hitting set $Y$ for some $\cH_{S_1,S_2}$ with $\alpha(Y)\leq d_{\ref{lem:separation}}$.
As \cref{lem:packing} was invoked in $G[W]$, we have that $Y \subseteq W$. Thus, by \cref{clm:hitting-to-separator}, $Y$ is an $S_1$--$S_2$-separator, and we can return $S_1,S_2,Y$ as the second outcome of \cref{lem:separation}.

Note that there are $\binom{q}{2}$ pairs $S_1,S_2$ and each family can be constructed by exhaustively enumerating paths of length at most $t$. As all parameters are fixed, the running time is polynomial.
\end{proof}

\subsection{Obtaining a balanced extended strip decomposition}

Bourneuf, Masa\v{r}\'ikov\'a, Nadara, and Pilipczuk~\cite{DBLP:conf/mfcs/BourneufMNP25} proved that, given an $S_{t,t,t}$-free graph and a mass function, one can remove a constant number of closed neighborhoods to obtain a rigid balanced extended strip decomposition of the remaining graph.

\begin{theorem}[{\cite[Theorem 3]{DBLP:conf/mfcs/BourneufMNP25}}]\label{thm:BMNP}
Given an $S_{t,t,t}$-free graph $G$ and a mass function $\mu$,
one can in polynomial time find $S\subseteq V(G)$ with $|S|\leq 3t+11$
and a rigid extended strip decomposition of $G-N_G[S]$ that is balanced for $\mu$.
\end{theorem}

In this subsection we aim to obtain an analogous statement, but instead of removing few closed neighborhoods,
we will remove a set of vertices of bounded independence number.
We remark that an existential statement of \cref{lem:alpha-esd} can be quite easily obtained by combining \cref{thm:BMNP} with a result of Hajebi and Spirkl~\cite[Lemma 7.2]{DBLP:journals/corr/abs-2605-01223}. However, we need an algorithmic variant.

\begin{lemma}\label{lem:alpha-esd}
For fixed $s,t$, there is a computable constant $a_{\ref{lem:alpha-esd}}$ and a polynomial-time
algorithm which, given an $n$-vertex $(S_{t,t,t},K_{s,s})$-free graph $G$ with mass function $\mu$,
finds $Z$ with $\alpha(Z)\leq a_{\ref{lem:alpha-esd}}$ and an extended strip decomposition
of $G-Z$ that is balanced for $\mu$ and whose host has $\Oh(n)$ vertices and edges.
\end{lemma}
\begin{proof}
Define the following constants depending only on $s,t$:
\begin{align*}
    b &=3t+11,\\
    c &=c_{\ref{lem:separation}}(b,s,t),\\
    d &=d_{\ref{lem:separation}}(b,s,t),\\
    a_{\ref{lem:alpha-esd}}&=2c+d.    
\end{align*}
Throughout the execution of the algorithm, we will maintain a $b$-system
$\cS$; the algorithm starts with $\cS=\emptyset$.
With each $S\in\cS$ we store a set $X_S$ with
$\alpha(X_S)\leq c$ and a rigid extended strip decomposition $(H_S,\eta_S)$ of
$G-(X_S\cup N_G[S])$ that is balanced for $\mu$.
By \cref{lem:restriction}\ref{it:restriction-rigid}, every stored host has at most $2n$ vertices and $n$ edges.

Iteratively, we apply \cref{lem:separation} to the current system $\cS$.

Suppose first that the application returns the second outcome, i.e., two anticomplete members $S_1,S_2\in\cS$ and an $S_1$--$S_2$-separator $Y$ with $\alpha(Y)\leq d$.
Define $Z=X_{S_1}\cup X_{S_2}\cup Y$, so $\alpha(Z)\leq2c+d=a_{\ref{lem:alpha-esd}}$.
No component of $G-Z$ meets both $N_G[S_1]$ and $N_G[S_2]$:
such a component, together with an endpoint in each $S_i$, would give
an $S_1$--$S_2$-path in $G-Y$.
Let $V_1$ be the union of the components avoiding $N_G[S_1]$, and set
$V_2=V(G-Z)\setminus V_1$. Then $V_1,V_2$ are anticomplete and
\[
 V_i\subseteq V\bigl(G-(X_{S_i}\cup N_G[S_i])\bigr)\qquad \text{ for } i \in \{1,2\}.
\]
Restrict $(H_{S_i},\eta_{S_i})$ to $V_i$ and take the disjoint union of
the two decompositions. By \cref{lem:restriction}\ref{it:restriction-induced} and~\ref{it:restriction-union}, this is a decomposition
of $G-Z$, and each new particle is contained in a stored particle.
Hence the result is balanced for $\mu$. Restriction leaves the hosts
unchanged, so their union has at most $4n$ vertices and at most $2n$ edges.
Return $Z$ and this decomposition, and terminate.
We remark that the restrictions may destroy rigidity, but the outcome of this step is never stored in $\cS$.

So, suppose that the application of \cref{lem:separation} returns the first outcome, i.e., a hitting set $X$ for $\cS$ with $\alpha(X)\leq c$.
Apply \cref{thm:BMNP} to $G-X$ with the restriction of $\mu$.
We obtain a set $S\subseteq V(G)\setminus X$ with $|S|\leq b$ and a rigid extended strip decomposition $(H,\eta)$ of
$(G-X)-N_{G-X}[S]=G-(X\cup N_G[S])$ whose particles have mass at most
$\mu(G-X)/2\leq\mu(G)/2$.
If $S=\emptyset$, return $Z=X$ and $(H,\eta)$, and terminate.
This satisfies the required independence-number bound since $\alpha(X)\leq c\leq a_{\ref{lem:alpha-esd}}$,
and the host has at most $2n$ vertices and $n$ edges by rigidity.

Otherwise, add $S$ to $\cS$, set $X_S=X$ and $(H_S,\eta_S)=(H,\eta)$, and proceed to the next iteration. Note that
$S$ was not already a member of $\cS$: it is disjoint from $X$, whereas every previous member meets $X$.
Thus every nonterminating iteration adds a new member to $\cS$, and there are at most $n^b$ possible members.
Each iteration takes polynomial time, using \cref{lem:separation,thm:BMNP} and, in the terminating separator case, restriction and disjoint union.
Hence the entire algorithm runs in polynomial time.
\end{proof}

\subsection{Small interfaces and component boundaries}

Fix $s\geq2$ and let $(H,\eta)$ be an extended strip decomposition of a $K_{s,s}$-free graph $G$.
Call a vertex set \emph{small} if its independence number is at most $s-1$, and \emph{large} otherwise.
At most one interface at a given host vertex is large:
two such interfaces would contain independent $s$-vertex sets complete to one another, inducing a $K_{s,s}$.
For every $x\in V(H)$, define
\begin{equation}\label{eq:small}
 \small(x)=\bigcup_{\substack{xy\in E(H)\\\eta(xy,x)\text{ small}}}\eta(xy,x),
 \qquad \Small=\bigcup_{x\in V(H)}\small(x).
\end{equation}
Thus $\Small$ is the union of all small interfaces.
As all interfaces at $x$ are complete to each other, we have $\alpha(\small(x))\leq s-1$.
However, the independence number of $\Small$ might be arbitrarily large.
For fixed $s$, sets $\small(x)$ and $\Small$ can be found in polynomial time by exhaustively enumerating $s$-vertex sets in each interface.

\begin{lemma}\label{lem:components}
Let $s\geq2$ be an integer, let $(H,\eta)$ be an extended strip decomposition of a $K_{s,s}$-free graph $G$.
Let $\small(x)$ for each $x \in V(H)$ and $\Small$ be defined as in \eqref{eq:small}.
Then the following hold.
\begin{enumerate}[label=(\arabic*),ref=(\arabic*)]
    \item\label{it:components-particles} $\alpha(\Small\cap A)\leq2(s-1)$ for every particle $A$ of $(H,\eta)$.
    \item\label{it:components-boundary} Every component $D$ of $G-\Small$ is contained in a particle of $(H,\eta)$ and satisfies $\alpha(N_G(D))\leq2(s-1)$.
\end{enumerate}
\end{lemma}
\begin{proof}
The set $\Small$ lies entirely in the union of edge atoms, and, for every edge $xy \in E(H)$, $\Small\cap\eta(xy)$ is the union of the small interfaces of $xy$.
Hence $\alpha(\Small\cap\eta(xy))\leq2(s-1)$.
Since every particle meets at most one edge atom, this proves item~\ref{it:components-particles}.

To prove item~\ref{it:components-boundary}, let $D$ be a component of $G-\Small$; in particular, $N_G(D)\subseteq\Small$.

\begin{claim}\label{clm:components-particle}
The component $D$ is contained in a particle of $(H,\eta)$.
\end{claim}
\begin{claimproof}
Let $(H,\eta')$ be the extended strip decomposition of $G-\Small$ obtained by restricting $(H,\eta)$ to $V(G)\setminus\Small$, as in \cref{lem:restriction}\ref{it:restriction-induced}.
Every small interface becomes empty, so at each host vertex at most one interface of $(H,\eta')$ is nonempty.
By the definition of an extended strip decomposition, every edge between two distinct edge atoms joins two interfaces at a common host vertex.
Hence distinct edge atoms of $(H,\eta')$ are anticomplete.
Similarly, all neighbors of a vertex atom $\eta'(x)$ lie in the unique nonempty interface at $x$, if it exists, so $\eta'(x)$ has neighbors in at most one edge atom.
A triangle atom $\eta'(xyz)$ has neighbors only in the edge atoms $\eta'(xy)$, $\eta'(xz)$, and $\eta'(yz)$.
Moreover, it has a neighbor in $\eta'(xy)$ only if both $\eta'(xy,x)$ and $\eta'(xy,y)$ are nonempty, and analogously for $xz$ and $yz$.
As any two of the edges $xy$, $xz$, and $yz$ of $H$ share a vertex, and at each host vertex at most one interface of $(H,\eta')$ is nonempty, at most one of these three edges has both of its interfaces in $(H,\eta')$ nonempty.
Hence $\eta'(xyz)$ also has neighbors in at most one edge atom.
Distinct vertex and triangle atoms are anticomplete.
Consequently, as $D$ is connected, it meets at most one edge atom of $(H,\eta')$.
If $D$ meets $\eta'(xy)$, then, as $D$ is connected and meets no other edge atom, every vertex or triangle atom meeting $D$ has a neighbor in $\eta'(xy)$.
Thus every such atom is $\eta'(x)$, $\eta'(y)$, or $\eta'(xyz)$ for some $xyz\in T(H)$, and hence $D\subseteq A_{xy}^{xy}$.
Otherwise, $D$ meets no edge atom, so it is contained in a single vertex or triangle atom, as distinct vertex and triangle atoms are anticomplete.
In both cases, $D$ is contained in a particle of $(H,\eta)$.
\end{claimproof}

Thus, it remains to show that $\alpha(N_G(D))\leq2(s-1)$.
Let us first consider the case when $D$ meets an edge atom $\eta(xy)$ for some $xy\in E(H)$.

\begin{claim}\label{clm:components-edge}
If $D$ meets $\eta(xy)$ for some $xy\in E(H)$, then $N_G(D)\subseteq\small(x)\cup\small(y)$.
In particular, $\alpha(N_G(D))\leq\alpha(\small(x))+\alpha(\small(y))\leq2(s-1)$.
\end{claim}
\begin{claimproof}
Every particle meeting $\eta(xy)$ is one of $A_{xy}^{\perp}$, $A_{xy}^{x}$, $A_{xy}^{y}$, and $A_{xy}^{xy}$, and all of them are contained in $A_{xy}^{xy}$.
Hence $D\subseteq A_{xy}^{xy}$ by \cref{clm:components-particle}.
Neighbors of $D$ in $\eta(xy)$ lie in $\Small\cap\eta(xy)$, and thus in $\small(x)\cup\small(y)$.
Since $N_G(D)\subseteq\Small$, every neighbor of $D$ lies in an edge atom.
By the definition of an extended strip decomposition, every neighbor of $D$ outside $\eta(xy)$ thus lies in an interface $\eta(xz,x)$ with $z\in N_H(x)\setminus\{y\}$ or in an interface $\eta(yz,y)$ with $z\in N_H(y)\setminus\{x\}$.
By symmetry, let $w$ be a neighbor of $D$ of the first kind; we show that $w\in\small(x)$.
The only vertices of $A_{xy}^{xy}$ that can be adjacent to $w$ lie in $\eta(x)$, in $\eta(xy,x)$, or in $\eta(xyz)$ if $xyz\in T(H)$, so $D$ meets one of these sets.
In fact, $D$ meets $\eta(xy,x)$.
Indeed, all neighbors of $\eta(x)$ and of $\eta(xyz)$ in $A_{xy}^{xy}$ lie in $\eta(xy,x)$, and if $D$ meets one of these two atoms, then it also contains a neighbor of this atom, as $D$ is connected and meets $\eta(xy)$.
As $D\cap\Small=\emptyset$, the interface $\eta(xy,x)$ is not contained in $\Small$, so it is large.
Since at most one interface at $x$ is large, $\eta(xz,x)$ is small, and hence $w\in\small(x)$.
Now, the bound on $\alpha(N_G(D))$ follows from the bound on $\alpha(\small(x))$ and $\alpha(\small(y))$.
\end{claimproof}

Now, consider the case when $D$ meets no edge atom, i.e., it is contained in a single vertex or triangle atom.
Since $N_G(D)\subseteq\Small$, the definition of an extended strip decomposition implies the following:
\begin{itemize}
\item if $D\subseteq\eta(x)$ for some $x\in V(H)$, then $N_G(D)$ is contained in the union of the sets $\Small\cap\eta(xy,x)$ over all $y\in N_H(x)$;
\item if $D\subseteq\eta(xyz)$ for some $xyz\in T(H)$, then $N_G(D)$ is contained in the union of the sets $\Small\cap\eta(xy,x)\cap\eta(xy,y)$, $\Small\cap\eta(xz,x)\cap\eta(xz,z)$, and $\Small\cap\eta(yz,y)\cap\eta(yz,z)$.
\end{itemize}
In both cases, any two of these sets are contained in two distinct interfaces at a common host vertex, and hence they are complete to each other.
Thus every independent subset of $N_G(D)$ is contained in one of these sets.
Moreover, each of these sets is contained in $\Small\cap A$ for some full edge particle $A$, so its independence number is at most $2(s-1)$ by item~\ref{it:components-particles}.
Therefore $\alpha(N_G(D))\leq2(s-1)$, which completes the proof of item~\ref{it:components-boundary}.
\end{proof}

\section{Excluding a single long claw}

In this section we prove \cref{thm:main} for $F=S_{t,t,t}$, using \cref{lem:alpha-esd,lem:components}.
We follow the recursive scheme of Abrishami, Chudnovsky, Dibek, Pilipczuk, and Rz\k{a}\.{z}ewski~\cite[Section~3]{10.1145/3844498}, keeping terminal sets of bounded independence number.

Fix integers $t \geq 1,s \geq 2$; we treat them as fixed constants.
Fix an ambient $(S_{t,t,t},K_{s,s})$-free weighted graph $(G,\weight)$, which will remain unchanged throughout the recursion.

\subsection{Auxiliary problem}
In the recursion, we solve a more general problem than the original one.
An \emph{instance} is a pair $(V',T)$ of disjoint subsets of $V(G)$, and its graph is $G'=G[V'\cup T]$ with weights $\weight$.
The vertices of $T$ are called \emph{terminals} and the set $V'$ is called the \emph{interior} of the instance.

The goal for an instance $(V',T)$ is to compute the auxiliary table entries $\Tab[V',T,I]$ for all independent sets $I\subseteq T$.
Each entry stores the largest weight that the interior can contribute when precisely the terminals in $I$ are selected:
\[
 \Tab[V',T,I]=\max\{\weight(J)\mid J\subseteq V',\ I\cup J\text{ is independent in }G\}.
\]
In particular, $\Tab[V(G),\emptyset,\emptyset]=\alpha(G,\weight)$.

For a set $Z\subseteq V'\cup T$ and an independent set $J\subseteq T\cup Z$, define
$W_J=V'\setminus(Z\cup N_G(J))$. Then, for every independent $I\subseteq T$,
\begin{equation}\label{eq:conditioning}
 \Tab[V',T,I]=
 \max_{\substack{\text{independent } J\subseteq T\cup Z\\J\cap T=I}}
 \left[\weight(J\setminus T)+\alpha(G[W_J],\weight)\right].
\end{equation}
Indeed, after fixing the selected vertices $J$ of $T\cup Z$, the remaining set is an independent set in $G[W_J]$.

\subsection{Child instances}\label{sec:reconstruction}

Fix an instance $(V',T)$ with graph $G'=G[V'\cup T]$.
We define child instances whose tables determine $\Tab[V',T,\cdot]$.
Let $Z\subseteq V'\cup T$ and let $(H,\eta)$ be an extended strip decomposition of $G'-Z$.
Let $\Small$ be defined as in~\eqref{eq:small}, and define $S=T\cup Z\cup\Small$.
For every particle $A$, let $\cQ(A)$ consist of the vertex sets of the components of $G[A\setminus S]$.
Define $\cQ=\bigcup_A\cQ(A)$, where the union ranges over all particles of $(H,\eta)$.
For each $R\in\cQ$, define its \emph{boundary} $B_R=N_G(R)\cap S$ and the child instance $(R,B_R)$, whose interior is $R$.
Note that even when $R\in\cQ(A)$, its boundary $B_R$ need not be contained in $A$.
We compute each child table once, even if $R$ occurs for several particles.

Clearly, one can construct $\cQ$ and all child instances in time polynomial in $|V'\cup T|+|V(H)|$.

\begin{lemma}\label{lem:replication}
Every vertex of $V'$ belongs to at most four members of $\cQ$.
Consequently, we have
\[
 \sum_{R\in\cQ}|R|\leq4|V'|.
\]
\end{lemma}
\begin{proof}
Every particle is contained in $(V'\cup T)\setminus Z$ and $T\cup Z\subseteq S$, so every member of $\cQ$ is a subset of $V'\setminus S$.
Hence the second statement follows from the first, and it suffices to prove the first statement for a fixed vertex $v\in V'\setminus S$.
For every particle $A$ containing $v$, let $C_A$ be the vertex set of the component of $G[A\setminus S]$ containing $v$.
Then $C_A$ is the unique member of $\cQ(A)$ containing $v$, so the members of $\cQ$ containing $v$ are exactly the sets $C_A$, where $A$ ranges over the particles containing $v$.

If $v$ lies in an edge atom $\eta(xy)$, then every particle containing $v$ is one of $A_{xy}^{\perp}$, $A_{xy}^{x}$, $A_{xy}^{y}$, and $A_{xy}^{xy}$.
If $v$ lies in a triangle atom $\eta(xyz)$, then every particle containing $v$ is one of $A_{xyz}$, $A_{xy}^{xy}$, $A_{xz}^{xz}$, and $A_{yz}^{yz}$.
In both cases, there are at most four sets $C_A$.

It remains to consider the case that $v$ lies in a vertex atom $\eta(x)$ for some $x\in V(H)$.
By the definition of particles, the particles containing $v$ are exactly the following:
\begin{itemize}
\item the vertex particle $A_x$, and
\item the half-edge particle $A_{xy}^{x}$ and the full edge particle $A_{xy}^{xy}$ for each $y\in N_H(x)$.
\end{itemize}
We show that at most three of the sets $C_A$ are distinct.

By the definition of an extended strip decomposition, every vertex of $N_{G'-Z}(\eta(x))$ lies in an interface $\eta(xy,x)$ for some $y\in N_H(x)$.
Let $y\in N_H(x)$ be such that $\eta(xy,x)$ is small, and let $A\in\{A_{xy}^{x},A_{xy}^{xy}\}$.
Since $A$ is disjoint from all edge atoms other than $\eta(xy)$, we have $N_{G'-Z}(\eta(x))\cap A\subseteq\eta(xy,x)\subseteq\Small\subseteq S$ by~\eqref{eq:small}.
Thus $\eta(x)$ is anticomplete to $A\setminus(S\cup\eta(x))$, and since $C_A$ is connected and contains $v\in\eta(x)$, we obtain $C_A\subseteq\eta(x)$.
As $A_x=\eta(x)\subseteq A$, this yields $C_A=C_{A_x}$.

Finally, recall that at most one interface at $x$ is large.
Hence every set $C_A$ equals $C_{A_x}$, except possibly $C_{A_{xy}^{x}}$ and $C_{A_{xy}^{xy}}$ for the unique $y\in N_H(x)$ with $\eta(xy,x)$ large, if such a $y$ exists.
Thus at most three of the sets $C_A$ are distinct.
\end{proof}

\begin{lemma}\label{lem:join}
Given the child tables, we can compute $\Tab[V',T,\cdot]$ in time polynomial in $|V'\cup T|+|V(H)|$ per independent subset of $T\cup Z$.
\end{lemma}
\begin{proof}
Fix an independent set $J\subseteq T\cup Z$ and let $W_J$ be as in~\eqref{eq:conditioning}.
For every particle $A$, we have
\begin{equation}\label{eq:join}
 \alpha(A\cap W_J,\weight)=\max_{\substack{I\subseteq\Small\cap A\cap W_J\\I\text{ independent}}}
 \left[\weight(I)+\sum_{R\in\cQ(A)}\Tab\bigl[R,B_R,(I\cup J)\cap B_R\bigr]\right].
\end{equation}
Indeed, after fixing $I$, the remaining choices lie in the pairwise anticomplete components $R\in\cQ(A)$.
As $I\cup J\subseteq S$, the choices in each $R$ must be compatible precisely with $(I\cup J)\cap B_R$, as encoded by the corresponding table entry.
By \cref{lem:components}\ref{it:components-particles}, there are $|V'\cup T|^{\Oh(s)}$ choices of $I$.
Restrict $(H,\eta)$ to $W_J$ using \cref{lem:restriction}\ref{it:restriction-induced}, and apply \cref{thm:assembly} with the values in~\eqref{eq:join} to compute $\alpha(G[W_J],\weight)$.
Equation~\eqref{eq:conditioning} gives the parent table.
This uses only the supplied child tables.
\end{proof}

\subsection{Proof of \cref{thm:main} for a single long claw}

\begin{theorem}[\cref{thm:main} for a single long claw]\label{thm:main-d1}
    For every fixed $s,t \geq 1$, there is a polynomial-time algorithm which, given an $(S_{t,t,t},K_{s,s})$-free graph $G$ and a weight function $\weight:V(G)\to\mathbb{R}_{\geq 0}$, returns a maximum-weight independent set in $(G,\weight)$.
\end{theorem}
\begin{proof}
Let $n=|V(G)|$.
If $n\leq1$ or $s=1$, the result is obvious.
Thus suppose $n\geq2$ and $s\geq2$, and define $k=a_{\ref{lem:alpha-esd}}+2(s-1)$; this constant depends only on $s,t$.

Throughout the recursion, we compute the table $\Tab[V',T,\cdot]$ for an instance $(V',T)$ while maintaining $\alpha(T)\leq6k$.
The initial call uses $V'=V(G)$ and $T=\emptyset$.
At each call, define $G'=G[V'\cup T]$.
If $|V'|\leq1$, compute each table entry directly.
Otherwise proceed as follows.
\begin{enumerate}
\item Enumerate all independent subsets of $T$ of size at most $6k$.
Choose one of maximum cardinality, say $J$, and write $h=|J|=\alpha(T)$.
\item If $h\leq5k$, let $\mu$ be the indicator mass function of $V'$.
Otherwise, let $\mu$ be the indicator mass function of $J$.
Apply \cref{lem:alpha-esd} to $G'$ with this mass function, obtaining $Z$ with $\alpha(Z)\leq a_{\ref{lem:alpha-esd}}$ and a balanced extended strip decomposition of $G'-Z$ with a host of linear size.
Define its small-interface set $\Small$ as in~\eqref{eq:small}.
\item Construct the distinct children $(R,B_R)$ as in \cref{sec:reconstruction} and recursively compute $\Tab[R,B_R,\cdot]$, once per child.
\item Compute the parent table $\Tab[V',T,\cdot]$ using \cref{lem:join}.
\end{enumerate}

We verify that the terminal bound is preserved and that the recursion terminates.
For a child $(R,B_R)$, let $D$ be the component of $G'-(Z\cup\Small)$ containing $R$.
Since $N_{G'}(D)\subseteq Z\cup N_{G'-Z}(D)$, \cref{lem:components}\ref{it:components-boundary} gives
\[
 \alpha(N_{G'}(D))\leq k,\qquad B_R\subseteq(T\cap D)\cup N_{G'}(D).
\]
If $h\leq5k$, the decomposition is balanced for the indicator mass function of the interior.
Since $D$ is contained in a particle, we obtain
\[
 |R|\leq|D\setminus T|\leq|V'|/2,\qquad
 \alpha(B_R)\leq\alpha(T\cap D)+k\leq h+k\leq6k.
\]

Suppose instead that $h>5k$.
The decomposition is balanced for the indicator mass function of $J$, so $|J\cap D|\leq h/2$.
Replacing $J\cap N_{G'}[D]$ by any independent set in $T\cap D$ preserves independence, as $J\setminus N_{G'}[D]$ is anticomplete to $D$.
Since $J$ is a maximum independent set of $T$, this gives
\[
 \alpha(B_R)\leq\alpha(T\cap D)+k
 \leq|J\cap D|+|J\cap N_{G'}(D)|+k
 \leq h/2+2k\leq5k.
\]
Thus the terminal bound is preserved, and any call that does not halve the interior has children with terminal independence number at most $5k$.
Every two recursion levels therefore halve the interior, so the recursion terminates.
Each child graph is induced in $G$ and hence remains $(S_{t,t,t},K_{s,s})$-free.
Correctness now follows by induction using \cref{lem:join}.
At the root, $\Tab[V(G),\emptyset,\emptyset]=\alpha(G,\weight)$.

The recursion has depth $\Oh(\log n)$, and \cref{lem:replication} bounds the sum of the interior sizes at depth $\ell$ by $4^\ell n$.
Since all child interiors are nonempty, the recursion tree has polynomially many nodes.
At each node, \cref{lem:alpha-esd} takes polynomial time and produces a host of linear size.
The bounds $\alpha(T)\leq6k$ and $\alpha(T\cup Z)\leq7k$ give polynomially many table entries and assignments in \cref{lem:join}.
Hence each node takes polynomial time apart from its recursive calls, and the total running time is polynomial.
Finally, a maximum-weight independent set can be recovered by standard bookkeeping.
\end{proof}


\section{Excluding a disjoint union of long claws}

In this section we prove \cref{thm:main}.
On a high level, another application of \cref{lem:packing} reduces the problem to the case of a single long claw, which can then be solved using \cref{thm:main-d1}.

\thmmain*
\begin{proof}
Let $d$ be the number of components of $F$ and let $t=\max\{1,|V(F)|\}$.
Then $F$ is an induced subgraph of $dS_{t,t,t}$, and hence every $(F,K_{s,s})$-free graph is also $(dS_{t,t,t},K_{s,s})$-free.

Define $b=3t+1$, and let $\cS$ consist of the vertex sets of all induced copies of $S_{t,t,t}$ in $G$.
This is a $b$-system that can be constructed in polynomial time by enumerating all $b$-vertex subsets of $V(G)$.
Apply \cref{lem:packing} to $\cS$ with parameters $b,s,d$.
The second outcome gives $d$ pairwise anticomplete induced copies of $S_{t,t,t}$, inducing $dS_{t,t,t}$ in $G$.
Thus we obtain a hitting set $X$ with $\alpha(X)\leq c_{\ref{lem:packing}}(b,s,d)$.
Since $X$ meets every induced copy of $S_{t,t,t}$, the graph $G-X$ is $S_{t,t,t}$-free.

Clearly,
\[
 \alpha(G,\weight)=
 \max_{\substack{J\subseteq X\\J\text{ independent}}}
 \left[\weight(J)+\alpha\bigl(G-(X\cup N_G(J)),\weight\bigr)\right].
\]
The bound on $\alpha(X)$ lets us enumerate all independent subsets $J\subseteq X$ in polynomial time.
For each $J$, the residual graph is induced in $G-X$ and hence is $(S_{t,t,t},K_{s,s})$-free.
Solve \MWIS in this graph using \cref{thm:main-d1}, and add $J$ to the resulting independent set.
Returning the heaviest set obtained in this way gives a maximum-weight independent set in $G$.
Since $d,s,t$ are fixed, the construction of $\cS$, the application of \cref{lem:packing}, and all residual computations take polynomial time.
\end{proof}


\section{Conclusion}

The main challenge remains to solve \MWIS in polynomial time in graphs excluding an arbitrary fixed subdivided claw, with no further assumptions.
We believe that extended strip decompositions are the right tool, but that further progress will require controlling the structure of particles, rather than just their size.
The goal would be to find, in polynomial time, a decomposition whose particles admit efficient algorithms for \MWIS.
The structure theorem of Chudnovsky and Seymour for claw-free graphs~\cite{DBLP:journals/jct/ChudnovskyS08e} provides a model for such an approach.

Abrishami et al.~\cite{ACDPR} considered $F$-free graphs, for $F\in\cF$, excluding a fixed biclique as a subgraph.
By Ramsey's theorem, this restriction is equivalent, up to changing the parameters, to excluding both a fixed clique and a fixed biclique as induced subgraphs.
Our result strengthens this result by removing the clique exclusion.
The complementary direction is to retain only the clique exclusion.
This direction was proposed by Abrishami et al.~\cite[Conjecture~10]{ACDPR}, and we restate it here.

\begin{problem}
    For every fixed $F\in\cF$ and integer $s\geq 1$, can \MWIS be solved in polynomial time in $(F,K_s)$-free graphs?
\end{problem}

Even the special case where $F$ is a path remains open in general.

\begin{problem}
    For every fixed pair of integers $t,s\geq 1$, can \MWIS be solved in polynomial time in $(P_t,K_s)$-free graphs?
\end{problem}

This question (in the unweighted setting) was explicitly asked by Chudnovsky, King, Pilipczuk, Rz\k{a}\.{z}ewski, and Spirkl~\cite{DBLP:journals/siamdm/ChudnovskyKPRS21}, and remains wide open.
Recall from the introduction that recently the case $t=7$ was settled in the affirmative~\cite{chudnovsky_et_al:LIPIcs.ISAAC.2025.20}.

\paragraph{Acknowledgement.} ChatGPT was used to simplify some constructions and to identify or correct omissions and minor mistakes and Claude was used to polish the write-up. The authors take full responsibility for the content of the paper.

\bibliographystyle{abbrv}
\bibliography{main}
\end{document}